\documentclass[
 superscriptaddress,
nofootinbib,
 amsmath,amssymb,
 twocolumn,
 aps,
 prd,
longbibliography
]{revtex4-2}

\usepackage[utf8]{inputenc}
\usepackage{geometry}
\usepackage{float}
\usepackage{graphicx,xcolor}
\usepackage{enumitem}
\usepackage{braket}
\usepackage[colorlinks=true]{hyperref}
\usepackage{amsfonts,amssymb}
\usepackage{amsmath, mathtools}
\usepackage{caption,subcaption}

\usepackage[normalem]{ulem}

\usepackage{tikz}
\usetikzlibrary{arrows.meta}

\usetikzlibrary{positioning}
\usepackage[edges]{forest}
\usepackage{enumitem}
\usepackage{braket}
\usepackage{hyperref}
\usepackage{amsfonts,amssymb}
\usepackage{amsmath, mathtools}
\usepackage{caption,subcaption}
\usepackage{comment}

\usepackage{amsthm}

\theoremstyle{definition}
\newtheorem{definition}{Definition}

\begin{document}

\title{
On the reliability and accessibility of quantum measurement apparatuses}
\author{Nicola Pranzini}
\email{nicola.pranzini@helsinki.fi}
\address{Department of Physics, P.O.Box 64, FIN-00014 University of Helsinki, Finland}
\address{QTF Centre of Excellence, Department of Physics, University of Helsinki, P.O. Box 43, FI-00014 Helsinki, Finland}
\address{InstituteQ - the Finnish Quantum Institute, Finland}
\author{Paola Verrucchi}
\email{verrucchi@fi.infn.it}
\address{ISC-CNR, UOS Dipartimento di Fisica, Universit\`a di Firenze, I-50019, Sesto Fiorentino (FI), Italy}
\address{Dipartimento di Fisica e Astronomia, Universit\`a di Firenze, I-50019, Sesto Fiorentino (FI), Italy}
\address{INFN, Sezione di Firenze, I-50019, Sesto Fiorentino (FI), Italy}

\begin{abstract}
    We propose a classification of measurement apparatuses based on their reliability and accessibility. Our notion of reliability parameterises the possibility of getting unexpected wrong results when using the apparatus in a given time window, and the one of accessibility describes the energy cost required to make the apparatus interact with a measured system. The classification is obtained by relating an apparatus's reliability and accessibility to the time dependence of the overlap of its pointer states. As an example, we study a one-to-all qubit interaction in which all the qubits act as a measurement apparatus for the one. This model shows that using randomly selected couplings results in accessible but unpredictable measurement apparatuses. Conversely, apparatuses with uniform coupling exhibit higher reliability but are energetically more costly.
\end{abstract}

\maketitle

\section{Introduction}
Measurements are at the foundation of quantum theory: they appear as one of the postulates of Quantum Mechanics (QM) and play a crucial role in distinguishing classical and quantum physics~\cite{NielsenC10}. Indeed, QM is the first - and only - theory containing a description of the procedure experimenters use to extract information about the physical world and, since \textit{obtaining knowledge is the very reason for making a measurement, [...] formulating quantum mechanics and, more generally, quantum phenomena in terms of information should throw a new light on the problem of measurement}~\cite{Zurek_Congress90}. In other words, we should use information and measurements to unravel the true meaning of QM, which lies hidden under the unresolved issues presented by the theory. These are the measurement problem~\cite{Schlosshauer05} and an unambiguous separation between classical and quantum systems~\cite{Atmanspacher97}. Although connected, the two conundrums are of different nature: while the first deals with the experimenter's everyday experience that a state collapses after a measurement and might require supplementing QM with some additional postulate~\cite{Maudlin95}, the second can be solved within the standard postulates of quantum mechanics, at least in the case of effectively classical apparatuses measuring quantum systems~\cite{Zurek03}. Indeed, a clear cut between measurement apparatuses and measured systems can be enforced by analysing the information transfer in the measurement process \textit{before} any output is produced~\cite{vonNeumann55}, hence allowing a system-apparatus distinction without the need of entering into the (heated) discussion about the measurement problem. 

In practice, the distinction is made by realizing a (von Neumann) premeasurement by a unitary evolution coupling the measurement apparatus and the measured system, hence characterizing the two via the information transfer caused by the interaction~\cite{WheelerEtAl73,Zurek83}. Specifically, this interaction can be modelled via Ozawa's prescription~\cite{Ozawa84}. As it is clear, this procedure does not need a description of the measurement-induced state collapse, yet it provides a clear cut between measured systems and apparatuses.

Once measured systems and measuring apparatuses are identified, one can ask which kind of quantum systems and interactions make good measurement apparatuses: this is the task we accomplish in this work. To do so, we first need to introduce a notion of quality for an apparatus, which we provide in terms of its reliability and accessibility. These formally define when and to which extent a quantum system can be used as an apparatus, for a given observer. In particular, reliability relates to the possibility of getting unexpected wrong results when using the apparatus in a given time window selected by the experimenter, and accessibility parametrises the feasibility by the observer of using the apparatus, given a finite amount of energy they have at their disposal and the energy cost required to make the apparatus interact with the system to be measured. Therefore, this work aims to build a distinction between measured systems and measurement apparatuses via the Ozawa realisation of the von Neumann premeasurement, and then classify measurement apparatuses by their reliability and accessibility in an operational picture. This task is performed under very general conditions, and an explicit example is provided to clarify the procedure further.

The article is structured as follows. In Sec~\ref{s.systems and apparatuses}, we briefly recollect the ingredients needed for our discussion: decoherence, POVMs, and, more importantly, the Probability Reproducibility Condition (PRC). Then, we show how to physically implement von Neumann's premeasurement via the so-called Ozawa interaction. In Sec.~\ref{s.WPRC_&_Energy}, this construction is used to relate the PRC to decoherence. Then, we show that requiring both unitarity and the PRC causes a lack of time intervals to perform measurements and introduce a weak version of PRC for which the sets of times satisfying the new condition have Lebesgue measure greater than zero. This leads to a classification of measurement apparatuses based on their reliability (in a given time window). Moreover, a brief analysis of the typical energies needed to perform a measurement leads to a discussion about the energy and size constraints one should impose on physically accessible apparatuses. As an application, in Sec.~\ref{s.example}, we review Zurek's model of the environment-induced decoherence for a one-to-all interaction amongst qubits and show that this can be interpreted as a premeasurement interaction between a measured system (the one) and a measurement apparatus (the all). We use the example to show that random system-apparatus couplings give more accessible measurement apparatus, while all identical system-apparatus couplings provide better reliability. Finally, we give our conclusions in Sec.~\ref{s.conclusions}.

\section{Systems and measurements}
\label{s.systems and apparatuses}
In this section, we provide a concise overview of the tools of the theory of quantum measurement used in the article. In particular, we review the concept of the Probability Reproducibility Condition, which will play a crucial role in our later analyses. 

\subsection{Systems}
We consider an isolated bipartite system $\Psi$ composed of a principal system $\Gamma$ and an environment $\Xi$. We require the Hilbert space of $\Psi$ to factor into $\mathcal{H}_{\Gamma}$ and $\mathcal{H}_{\Xi}$ such that 
\begin{equation}
\text{dim}(\mathcal{H}_{\Gamma})=d \ll N = \text{dim}(\mathcal{H}_{\Xi})~,
\end{equation}
meaning that the environment is larger than the principal system. We always assume this condition holds when the former is a measuring apparatus testing the latter. The system's states are either represented by normalized vectors in $\mathcal{H}_\Psi$ or density operators in $\mathcal{H}_\Psi^*$~\cite{NielsenC10}, for which the state of one part is
\begin{equation}
\rho_{I}=\text{Tr}_{\Psi\cap I}[\rho_{\Psi}]~,~~I=\Gamma, \Xi~.
\label{e.partial trace}
\end{equation}

The time evolution of subsystems' states is then easily obtained as
\begin{equation}
\rho_{I}(t)=\text{Tr}_{\Psi\cap I}[\rho_{\Psi}(t)]=\text{Tr}_{\Psi\cap I}[\hat{\mathcal{U}}_{t,t_0}\rho_\Psi(t_0)\hat{\mathcal{U}}_{t,t_0}^{\dagger}]~,
\end{equation}
where $\hat{\mathcal{U}}_{t,t_0}$ is the global time evolution operator from $t_0$ to $t$. A type of evolution playing a relevant role in the following is decoherence, i.e. one leaving invariant the diagonal elements of the density operators and changing the off-diagonal ones. As it is clear, specifying the basis with respect to which the decoherence occurs is necessary, as the notion of diagonal elements is basis-dependent. Hence, this dynamic is usually called \textit{decoherence with respect to a basis}~\cite{BreuerP02}. 

\subsection{Measurements}
\label{s.ozawa}

\subsubsection{POVMs and the Probability Reproducibility Condition}
Given a $\sigma$-algebra $\mathcal{M}$ obtained as the family of parts of a collection $\{E_{l}\}$ of all possible results of a given measurement, we define the associated \textit{positive operator-valued measure} (POVM) as the application
\begin{equation}
O:\mathcal{M}\longrightarrow op(\mathcal{H})
\end{equation}
satisfying some standard conditions that can be found, for example, in Ref.s~\cite{NielsenC10, BreuerP02, wisemanM10}. Given this setup, we define a mapping from $\mathcal{M}$ to $[0,1]$, specified by 
\begin{equation}
    p^{O}_{\rho}(E_l)=\text{Tr}[\rho\hat{O}(E_l)]~,
\end{equation}
as the function giving the probability of an experimental observation associated with the measurement $O$ of the system in the state $\rho$ to give the result $E_l$~\cite{HeinosaariZiman11}.

Given the generality of the above picture, it is often useful to associate the concept of POVM measurement to that of measurement apparatuses. These are additional quantum systems able to implement POVMs through some physical procedure. In particular, one can add an apparatus to the picture by promoting $\Xi$ to be the measurement apparatus. Then, we assign a collection of measurement results $\{E_{\gamma}\}$ and $\{E_{\xi}\}$ to each subsystem of $\Psi$, from which we build the respective family of parts $\mathcal{M}_{\Gamma}$ and $\mathcal{M}_{\Xi}$. Measurements performed on $\Xi$ by means of
\begin{equation}
    O_{\Xi}:\mathcal{M}_{\Xi}\rightarrow op(\mathcal{H}_{\Xi})
\end{equation}
are often called \textit{pointers}~\cite{Zurek81,Zurek03}. Next, we define an invertible function $f:\mathcal{M}_{\Xi}\rightarrow\mathcal{M}_{\Gamma}$ and obtain  the so-called \textit{probability reproducibility condition} (PRC) as a crucial condition that must hold for an apparatus to be considered good~\cite{Lahti91,BuschEtAl96}, i.e.
\begin{equation}
p^{O_{\Gamma}}_{\rho_{\Gamma}}(E_{\gamma})=p^{O_{\Xi}}_{\rho_{\Xi}}(f^{-1}(E_{\gamma}))~;
\end{equation}
in words, this condition requires that probabilities associated with outcomes relative to experiments performed on $\Gamma$ are equal to those associated with outcomes relative to experiments performed on the apparatus $\Xi$.

The process of measuring a system using an apparatus is composed of the \textit{premeasurement} and 
the \textit{output production} (or \textit{measurement-induced state collapse})~\cite{Schlosshauer07}. The first is a dynamical process generating entanglement between and transferring probabilities from the measured system and the measuring apparatus, as established by the the PRC~\cite{vonNeumann55, BuschEtAl96}. Next, in the output production, a definite outcome is selected from the quantum superposition generated during the premeasurement. Notice that having a full description of the premeasurement does not imply a description of the output production, while the converse might be true~\cite{Pessoa97}. However, the output production remains a controversial topic and subject to intense debate, as different interpretations of quantum mechanics offer contrasting descriptions of it~\cite{Schlosshauer13}. Addressing this issue is commonly referred to as the measurement problem, and interpretations of quantum mechanics can be classified based on their proposed solutions to the issue~\cite{Maudlin95}. Fortunately, our following discussion will focus solely on the premeasurement, which we often call measurement for the sake of simplicity.

\subsubsection{von Neumann's premeasurement and Ozawa's Hamiltonian}
The premeasurement can be described by von Neumann's scheme~\cite{vonNeumann55}. Given a basis ${\ket{\gamma}}$ of $\mathcal{H}_{\Gamma}$ over which the systems state is read as
\begin{equation}
    \ket{\Gamma}=\sum_{\gamma} c_{\gamma}\ket{\gamma}~,
\end{equation}
and an initial 'ready-to-read' state of the apparatus $\ket{\Xi_R}$, von Neumann's premeasurement is defined by the mapping
\begin{equation} 
\left(\sum_{\gamma} c_{\gamma}\ket{\gamma}\right)\otimes\ket{\Xi_R}\rightarrow\sum_{\gamma}c_{\gamma}\ket{\gamma}\otimes\ket{\Xi^{\gamma}}~.
\label{e.vonNeumann} 
\end{equation}
In this context, the PRC reads
\begin{equation} 
\langle\Xi^{\gamma}\ket{\Xi^{\gamma'}}=\delta_{\gamma,\gamma'}
\label{e.PRC}
\end{equation}
\cite{BuschEtAl95, HeinosaariZiman11}. von Neumann's premeasurements present the so-called problem of the preferred basis~\cite{Giulini96, HemmoO22}: while the initial state can be described in any basis of $\mathcal{H}_\Gamma$, and all of them provide equally good descriptions of the system, choosing one in Eq.~\eqref{e.vonNeumann} determines the physical observable being measured via the vectors $\ket{\Xi^{\gamma}}$. Hence, using von Neumann's scheme alone does not provide enough details to select which quantities are being measured independently of the initial arbitrary basis choice. This problem can be solved by specifying a unitary Hamiltonian evolution realizing the premeasurement, called Ozawa's Hamiltonian~\cite{Ozawa84, Ozawa23}. To understand this, let us assume the system $\Psi$ evolves via the local Hamiltonian
\begin{equation}
\hat{H}=\hat{O}_{\Gamma}\otimes\hat{O}_{\Xi}~,
\label{e.ozawaH}
\end{equation}
where $\hat{O}_{\Gamma}$ is diagonalised by $\{\ket{\gamma}\}$ as
\begin{equation}
\hat{O}_{\Gamma}=\sum_{\gamma}\omega_{\gamma}\ket{\gamma}\bra{\gamma}~.
\end{equation}
Then, at time $t$ the initial separable state appearing on LHS of Eq.~\eqref{e.vonNeumann} becomes
\begin{equation}
\ket{\psi(t)}=\sum_{\gamma}c_{\gamma}\ket{\gamma}\otimes\ket{\Xi^{\gamma}(t)}~,
\label{Ozawa}
\end{equation}
where we defined
\begin{equation}
    \ket{\Xi^{\gamma}(t)}= e^{-i\omega_{\gamma}\hat{O}_{\Xi}t}\ket{\Xi_R}~.
    \label{e.xi_gamma}
\end{equation}
As it is clear, this state can exhibit the entanglement structure required by von Neumann's premeasurement without the need to initially specify the basis of $\Gamma$, as this is automatically provided by the form of the interaction \eqref{e.ozawaH}. Therefore, by selecting the basis for the von Neumann process, Ozawa's Hamiltonian removes the arbitrariness in the choice of decomposition \eqref{e.vonNeumann} and solves the problem of the preferred basis~\cite{Schlosshauer05,Schlosshauer07, Zurek81}. This way of solving the problem, called \textit{stability criterion}, is part of a larger program aiming at solving the quantum-to-classical crossover and the measurement problem via environment-induced decoherence~\cite{Zurek82,Zurek00,Riedel12,Zurek09,BlumeKohout05}. Moreover, Eq.s~(\ref{e.ozawaH}-\ref{Ozawa}) provide a physical description of a system-environment interaction we can use for designing measurement apparatuses. However, since using the states \eqref{e.xi_gamma} as pointers means that Eq.~\eqref{e.PRC} cannot hold at all times, a thorough classification of measurement apparatuses based on the range of times during which the PRC holds is necessary. This is the task we accomplish in the next section.

\section{Physical requirements on measurement apparatuses}
\label{s.WPRC_&_Energy}
In addition to the essential requirement of generating entanglement between the system and the apparatus prescribed by Eq.s~(\ref{e.vonNeumann}-\ref{e.PRC}), a physical measurement apparatus must also satisfy physically grounded demands regarding the length of the time windows for which the information about $\Gamma$ can be extracted from $\Xi$, and the finiteness of the energy scale introduced by the coupling between the two systems. In particular, we require the interaction to be such that 1) the time during which the PRC (or a weak version of it) is satisfied is long enough to allow an observer to manipulate the apparatus and extract information and 2) the interaction Hamiltonian is characterised by an energy scale large enough to not become irrelevant under noise and other perturbations and small enough to be physically accessible (e.g. the measurement process do not require infinite energy). We call these two requirements \textit{reliability} and \textit{accessibility} of a measurement apparatus; these are analysed in the subsections \ref{s.s.time} and \ref{s.s.energy}, respectively.

\subsection{Reliability}
\label{s.s.time}
Despite showing a formal resemblance with the RHS of ~\eqref{e.vonNeumann}, the state \eqref{Ozawa} is not always entangled. Indeed, whenever $\mathcal{H}_\Psi$ is finite-dimensional the system and apparatus are bounded to (quasi-)periodically return in the initial separable state. Moreover, not all times provide a faithful translation of the probabilities from the principal system to the apparatus; hence, a careful selection of the times the apparatus is read is required. To understand this, let us study the dynamics undergone by the reduced density operator for the principal system
\begin{equation}
    \rho_{\Gamma}(t)
    =\sum_{\gamma}|c_{\gamma}|^2\ket{\gamma}\bra{\gamma}+\sum_{\gamma,\gamma'}c_{\gamma}c_{\gamma'}^*\Delta_{\gamma'\gamma}(t)\ket{\gamma}\bra{\gamma'}~,
\end{equation}
where the function 
\begin{equation}
    \Delta_{\gamma'\gamma}(t)=\bra{\Xi^{\gamma'}(t)}\Xi^{\gamma}(t)\rangle
\end{equation}
describes a time dependence for the off-diagonal elements. As the density operator's diagonal elements remain constant, the Ozawa model describes decoherence for the principal system in the preferred basis, selected by the interaction. Similarly, the state of the apparatus at time $t$ is
\begin{equation}
    \rho_\Xi(t)=\sum_\gamma|c_\gamma|^2 \ket{\Xi^{\gamma}(t)}\bra{\Xi^{\gamma}(t)}~;
\end{equation}
by measuring on $\Xi$ the projectors 
\begin{equation}
    \Pi_\gamma=\ket{\Xi^{\gamma}(t)}\bra{\Xi^{\gamma}(t)}
\end{equation}
associated with the outcomes $f^{-1}(\gamma)$, we get the outcome $f^{-1}(\gamma')$ with probability
\begin{equation}
    p(f^{-1}(\gamma'))=\sum_\gamma |c_\gamma|^2 |\Delta_{\gamma'\gamma}(t)|^2~.
\end{equation}
Therefore, the PRC holds only at those times when
\begin{equation}
    |\Delta_{\gamma'\gamma}(t)|=\delta_{\gamma,\gamma'}
\end{equation}
and, at the same time, the off-diagonal elements of the principal system's density operator become zero: the PRC is realised only when the principal system fully decoheres~\cite{JoosZeh85}.

As a consequence of these considerations, we define the set $\mathbb{T}_{\rm{PRC}}$ of all times for which the PRC holds, i.e.
\begin{equation}
    \mathbb{T}_{\rm{PRC}}=\{t\in\mathbb{R}^+~s.t.~\lvert\Delta_{\gamma'\gamma}(t)\rvert=\delta_{\gamma\gamma'},~\forall \gamma,\gamma'\}~;
\end{equation}
as it is clear, this set plays a crucial role in establishing when an apparatus is good for its purpose. However, $\mathbb{T}_{\rm{PRC}}$ generally does not contain intervals. This is a problem because, for practical purposes, sufficiently long time intervals are needed for operating the apparatus and obtaining information about the system. To solve the issue, we introduce a weak version of the PRC (WPRC) by selecting a small parameter $\epsilon\geq0$ and defining the set $\mathbb{T}_{\rm{WPRC}}^{\epsilon}$ such that
\begin{equation}
\mathbb{T}_{\rm{WPRC}}^{\epsilon}=\{t\in\mathbb{R}^+~s.t.~\lvert\Delta_{\gamma'\gamma}(t)\rvert<\epsilon,~\forall \gamma\neq\gamma'\}~.
\label{e.WPRC}
\end{equation}
Hence, the pointer vectors become $\epsilon$-orthogonal at times contained in $\mathbb{T}_{\rm{WPRC}}^{\epsilon}$, which is a primary requirement for the emergence of definiteness in large-but-finite systems~\cite{CoppoEtAl22}. Clearly enough, $\mathbb{T}_{\rm{WPRC}}^{\epsilon}\to \mathbb{T}_{\rm{PRC}}$ in the limit of $\epsilon\to 0$, and the information extracted during $\mathbb{T}_{\rm{WPRC}}^{\epsilon}$ is smaller than that obtained at times in $\mathbb{T}_{\rm{PRC}}$ (see App.~\ref{s.mutual information} for further details).

With these definitions, we call \textit{reliable} a measurement apparatus for which it is possible to choose a small enough $\epsilon$ such that $\mathbb{T}_{\rm{WPRC}}^{\epsilon}$ contains sufficiently long intervals to operate over $\Xi$ while also keeping the WPRC as close as possible to PRC, meaning that the overlap between the pointer vectors is sufficiently small to make them distinguishable. When this happens, we can extract information about $\Gamma$ by measuring the pointer basis $\{\ket{\Xi^{\gamma}(t)}\}$ of (a subspace of) $\mathcal{H}_\Xi$. Physically, this means that an observer using $\Xi$ as apparatus and letting it interact with $\Gamma$ for doing measurements needs not to know the details of the interaction to get the correct results.

To make the idea above precise, let us consider an observer measuring $\Xi$ at times selected randomly within a time window $\mathbb{T}=[t_i,t_f]$. We call $G$ the collection of events for which they get a good measurement (PRC holds) and $B$ the one for which they get a bad measurement (PRC does not hold). Given this construction, we call a \textit{reliable measurement apparatus over} $\mathbb{T}$ one for which $G$ happens almost surely in $\mathbb{T}$, i.e. $P(G)=1$ when the apparatus is tested in $\mathbb{T}$. By construction, a reliable measurement apparatus over $\mathbb{T}$ is also reliable over all the intervals contained in $\mathbb{T}$. Moreover, we say one apparatus is more reliable than another if the measurement window of the first contains all those of the second. By extension, we call \textit{perfectly reliable} a measurement apparatus that is reliable over $\mathbb{R}^+$.

The above definitions can be related to the behaviour of $\Delta_{\gamma,\gamma'}(t)$ as follows. Let us consider an observer picking the times at which they measure following some reasonably well-behaved probability distribution $p(t)$. Given the measurement time window $\mathbb{T}$ of above, an appropriate probability distribution\footnote{For example, we can model an observer measuring within an interval centred in $t_m$ and having error over time estimation $\Delta t$ by the Gaussian distribution with $\mu=t_m$ and $\sigma=\Delta t/6$; in this case, we can select $\mathbb{T}=[t_m-\Delta t/2,t_m+\Delta t/2]$ to obtain $\int_\mathbb{T}p(t)\simeq 1$.} is one for which
\begin{equation}
    \int_{\mathbb{T}} p(t)dt=\Theta\simeq 1~.
\end{equation}
Then, given that we measure in $\mathbb{T}$, the probability of getting a good measurement is 
\begin{equation}
P(G)=\frac{1}{\Theta}\int_{\mathbb{T}\cap\mathbb{T}_{\rm{PRC}}}p(t)dt~;
\end{equation}
however, since for any finite-dimensional system the Lebesgue measure of $\mathbb{T}_{\rm{PRC}}$ is null, $P(G)=0$ and no good measurement apparatus (in the PRC sense) can be constructed. This is the ultimate reason why it is necessary to introduce the WPRC. Indeed, by using the latter in place of the former, we get that
\begin{equation}
    \int_{\mathbb{T}\cap\mathbb{T}^\epsilon_{\rm{WPRC}}}p(t)dt=\theta_\epsilon
    \label{e.good_or_perfect}
\end{equation}
is generally non-zero and, by comparing $\Theta$ and $\theta_\epsilon$, we obtain that
\begin{itemize}
    \item \textit{reliable} apparatuses over $\mathbb{T}$ satisfy $\theta_\epsilon=\Theta$;
    \item \textit{perfectly reliable} apparatuses satisfy $\theta_\epsilon=\Theta$ for any measurement window $\mathbb{T}$.
\end{itemize}
In particular, Eq.~\eqref{e.good_or_perfect} means that $\mathbb{T}\subseteq\mathbb{T}^{\epsilon}_{\rm{WPRC}}$ implies the apparatus is good for any observer measuring in time windows equal or smaller than $\mathbb{T}$. Notice that, as it is clear from the explicit $\epsilon$ dependence, the definitions of reliable and perfectly reliable apparatuses depend on the observer's precision requirements.

\subsection{Accessibility}
\label{s.s.energy}

Using Ozawa's Hamiltonian \eqref{e.ozawaH} hides the fact that any interaction must be turned on and off, requiring some energy for the switching~\cite{Abdelkhalek18}. To heuristically characterise this energy, let us suppose the spectrum of the interaction Hamiltonian is characterised by an energy scale $E_0$. Notice that this energy cost is only due to the interaction correlating $\Gamma$ and $\Xi$ and does not include the one required for the output production, which can be related to the average information extracted from the apparatus~\cite{DeffnerEtAl16,Jacobs12}.

On the one hand, for the interaction to be relevant the energy should be large enough to overcome any eventual environmental noise and external perturbations. For example, assuming that there is a fundamentally irreducible noise in the system whose typical energy scale is $\epsilon$, we must require
\begin{equation}
    E_0\gtrsim \epsilon~.
    \label{e.E small limit}
\end{equation}
On the other hand, considering that any observer only has limited resources to spend on building the interaction results in an upper limit for the usable interaction energy \cite{GuryanovaEtAl20}, such as
\begin{equation}
    E_0\lesssim \mathcal{E}~.
    \label{e.E big limit}
\end{equation}

As we will see in the next section, it is possible to schematise the premeasurement as an interaction between the measured system and a collection of subsystems composing the apparatus. In this case, we can assume each subsystem interacts independently with the measured one with a typical interaction energy scale $e_0$ and relate our considerations about the energy to the size of the apparatus. Then, the total energy of a measurement interaction needed to turn on and off an apparatus composed of $N$ subsystems is
\begin{equation}
    E_0\simeq Ne_0
\end{equation}
and the conditions \eqref{e.E small limit} and \eqref{e.E big limit} can be read as the lower and upper bounds
\begin{equation}
    \frac{\epsilon}{e_0}\simeq N_l< N < N_u\simeq\frac{\mathcal{E}}{e_0}
    \label{e.N_bound}
\end{equation}
over the possible number of subsystems used to build the apparatus. Therefore, we classify apparatuses based on their typical size as
\begin{itemize}
    \item \textit{nonfunctional}, if $N< N_l$;
    \item \textit{accessible}, if $N_l < N	< N_u$; 
    \item \textit{inaccessible}, if $N	> N_u$; 
\end{itemize}
It is important to notice that, while this classification depends on the chosen $\epsilon$, $\mathcal{E}$ and specific interaction considered, it describes a general and physically grounded picture that can be applied in most settings with little to none modification.

Given these definitions, the next section presents a concrete example of searching for reliable and accessible apparatuses over some given finite interval and shows that perfect apparatuses are inaccessible.

\section{Measuring a qubit via one-to-all interactions}
\label{s.example}

In this section, we study a qubit model realizing von Neumann premeasurement by the Ozawa scheme; the model was first proposed by Zurek in 1982~\cite{Zurek82} (see also Ref.s~\cite{Cucchietti05,Schlosshauer07}). After having presented the system and its general features, we discuss what makes a reliable and accessible apparatus in this case, in the sense of our classification presented in Sec.~\ref{s.WPRC_&_Energy}. In particular, we will define apparatuses interacting with the measured system in ordered and disordered ways, give a notion of quality of an apparatus, and compare ordered and disordered apparatuses to establish which choice is better between the two.

We consider a principal system made by one qubit (the measured system) and an apparatus composed of $N>1$ qubits, so that $dim(\mathcal{H}_{\Gamma})=2<dim(\mathcal{H}_{\Xi})=2^N$. Furthermore, we take the two systems to interact via the Ozawa-like interaction Hamiltonian
\begin{equation}
\hat{H}=\sigma_z\otimes\sum_{k=1}^N g_k \hat{S}_k~,
\label{e.H}
\end{equation}
with
\begin{equation}
\hat{S}_k=\underbrace{\mathbb{I}_2\otimes ... \otimes\mathbb{I}_2}_\text{k-1}\otimes\sigma_z\otimes\underbrace{\mathbb{I}_2\otimes ... \otimes\mathbb{I}_2}_\text{N-k}~.
\end{equation}
The scaling of the interaction energy as $E\sim \langle g\rangle N$, where $\langle g\rangle$ is the average of the couplings $g_k$, allows us to interpret the bounds discussed in Sec.~\ref{s.s.energy} as a direct constraint on the number of subsystems comprising the apparatus, as indicated by Eq.~\eqref{e.N_bound}. Therefore, we can employ $N$ to discriminate between nonfunctional, accessible, and inaccessible apparatuses. Moreover, let us suppose that, at time $t=0$, the system is in the separable state
\begin{equation}
\ket{\psi}=\left(a\ket{\uparrow}+b\ket{\downarrow}\right)\bigotimes_{k=1}^N\left(\alpha_k\ket{\uparrow_k}+\beta_k\ket{\downarrow_k}\right)~.
\end{equation}
Following the above description, at time $t$ the $N+1$ qubits will be in the state
\begin{equation}
    \ket{\psi(t)}=a\ket{\uparrow}\otimes\ket{\Xi^+(t)}+b\ket{\downarrow}\otimes\ket{\Xi^-(t)}~,
    \label{e.example_state}
\end{equation}
where we introduced the pointer states
\begin{equation}
    \ket{\Xi^\pm (t)}=\bigotimes_{k=1}^N\left(\alpha_k e^{\mp ig_kt}\ket{\uparrow_k}+\beta_k e^{\pm ig_kt}\ket{\downarrow_k}\right)~.
\end{equation}

This is the formal expression required by the von Neumann premeasurement scheme, but, as discussed in Sec.~\ref{s.ozawa}, a detailed analysis of the time dependence of the bipartite entanglement of the state \eqref{e.example_state} must be performed. As it is easy to show, the density operator of the measured qubit in the $\{\ket{\uparrow},\ket{\downarrow}\}$ basis is
\begin{equation}
\rho_{\Gamma}=\begin{pmatrix}
    |a|^2 & \gamma(t) \\
    \gamma^*(t)& |b|^2
\end{pmatrix}~,
\label{e.rho_gamma}
\end{equation}
where
\begin{equation}
\gamma(t)=ab^*\langle\Xi^-(t)\ket{\Xi^+(t)}
\end{equation}
and
\begin{equation}
    \langle\Xi^-(t)\ket{\Xi^+(t)}=\prod_{k=1}^N\left(|\alpha_k|^2e^{-2ig_kt}+|\beta_k|^2e^{2ig_kt}\right)
\label{cprod}
\end{equation}
Notice that the PRC holds iff $\langle\Xi^-(t)\ket{\Xi^+(t)}$ is null, and the WPRC holds iff its modulus is smaller than the $\epsilon$ appearing in the definition \eqref{e.WPRC}. For these reasons, we introduce the availability
\begin{equation}
    \mathcal{A}(t)=\lvert\langle\Xi^-(t)\ket{\Xi^+(t)}\rvert~,
\end{equation}
which is particularly informative for our purposes, as it captures when the PRC and WPRC are available (when the pointer states can be read as the time evolution under slightly different Hamiltonians, the above quantity is called Loschmidt echo~\cite{JalabertEtAl01, KarkuszewskiEtAl02, CucchiettiEtAl03, JafariJ17}).

It was shown by Zurek in Ref.~\cite{Zurek82} that the long-time average of $\mathcal{A}(t)$ goes to zero, and its oscillations are, on average, contained in a neighbourhood of zero bounded by the long-time variance
\begin{equation}
\text{Var}[\mathcal{A}(t)] =\prod_{k=1}^N\left(|\alpha_k|^4+|\beta_k|^4\right)~.
\end{equation}
From this expression, and assuming only a finite portion $M\ll N$ of the $\alpha_k$ to be $0$ or $1$ (which is reasonable whenever the apparatus has not been prepared \textit{ad hoc} to be in a specific state), we can find an $x$ such that
\begin{equation}
|\alpha_k|^4+|\beta_k|^4\leq x<1
\end{equation}
for all those $\alpha_k\neq 0,1$, from which follows that
\begin{equation}
    \text{Var}[\mathcal{A}(t)]\leq x^{N-M}\simeq 0
\end{equation}
for any sufficiently large apparatus. In this case, we can say that the overlap \eqref{cprod} is null almost everywhere, i.e. everywhere except from a discrete set of values of $t$.

\begin{figure}
     \centering
     \includegraphics[width=0.45\textwidth]{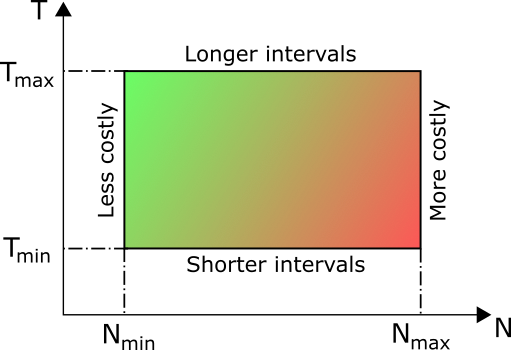}
    \caption{Classification of apparatuses based on the physical constraints relevant to the observer. The plot shows the number of subsystems comprising the apparatus vs. the duration of the provided measurement window. Apparatuses in the top-left corner of the box (green) are the best accessible ones, as they offer extended measurement windows at a lower energy cost; those in the bottom-right corner (red) are the worst accessible ones, generating smaller windows at a higher energy cost.}
    \label{f.plot}
\end{figure}

The dynamics of the measured qubit can be understood as the movement of a point within the Bloch-Majorana ball~\cite{Majorana32, Bloch46}. In this representation, any density operator of a single qubit can be expressed as
\begin{equation}
\rho_{\Gamma}=\frac{1}{2}(\mathbb{I}_2+\vec{r}\cdot\vec{\sigma})~,
\end{equation}
where $\vec{r}$ belongs to the three-dimensional ball $B^3$; pure and mixed states are respectively characterised by having $\lvert\vec{r}\rvert=1$ and $\lvert\vec{r}\rvert<1$. In particular, by Eq.~\eqref{e.rho_gamma}, we get that the point representing the qubit's state spends the majority of its time in a disk-shaped region at constant $z$ value, where $z = |a|^2 - |b|^2$, and whose radius is smaller for larger environments. For example, if we pick $a=b$, then it is
\begin{equation}
    |\vec{r}(t)|^2=\mathcal{A}(t)~,
\end{equation}
and the above statements clearly follow from the discussion about the long-time average and variance of $\mathcal{A}(t)$. Moreover, if we take $|\alpha_k|=|\beta_k|$, $\forall k$, we obtain
\begin{equation}
    \langle\Xi^-(t)\ket{\Xi^+(t)}=\prod_{k=1}^N\cos(2g_k t)~,
    \label{e.selected delta}
\end{equation}
meaning that the vector in the Bloch-Majorana ball starts its motion from a point on the unit radius circle at $z=0$ and evolves towards the inner region of the disk. At large $N$, $\mathcal{A}(t)$ rapidly approaches values close to zero and stays there for very long times, meaning that the decoherence is fast and persistent~\cite{JafariJ17}: the majority of time values will fall within the set $\mathbb{T}^\epsilon_{\rm{WPRC}}$. While this behaviour is generally observed regardless of the specific couplings involved, some distinctions exist between the ordered and disordered cases, defined as 
\begin{definition}[]\label{d.order}
A $g$-\textit{ordered apparatus} is an environment composed of $N$ qubits interacting with the principal system via \eqref{e.H} with $g_k=g,~\forall k$;
\end{definition}
and
\begin{definition}[]\label{d.disorder}
A $G_I$-\textit{disordered apparatus} is an environment composed of $N$ qubits interacting with the principal system via \eqref{e.H} with the $g_k$'s randomly sampled from a uniform distribution $G_{I}$, where $I$ is the sampling interval;
\end{definition} 
The differences between these apparatuses can be categorised according to the criteria discussed in Sec.~\ref{s.WPRC_&_Energy}, enabling a selection of the most suitable measurement apparatus for a given purpose. This is the task we accomplish hereafter.

\subsection{Ordered and disordered apparatuses}
\begin{figure}
     \centering
    \includegraphics[width=0.45\textwidth]{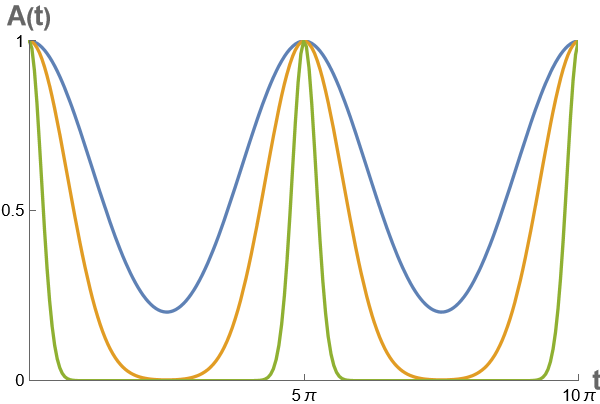}
    \caption{Availability $\mathcal{A}(t)$ for a $g$-ordered apparatus with $g=0.1$, for $N=5$ (blue), $N=10$ (orange), and $N=100$ (green).\vspace{1.2em}}
    \label{f.ordered}
\end{figure}

Before discussing which choice of couplings is the best, we must introduce a notion of quality for measurement apparatuses. This is done by the tools of Sec.~\ref{s.WPRC_&_Energy}, where we saw that apparatuses are characterized by the physical constraints describing the measurement capabilities of the observers operating them. Specifically, these are:
\begin{itemize}
    \item the duration $T$ of the time interval $\mathbb{T}^\epsilon_{\rm WPRC}$ provided to an observer to perform measurements (see Sec.~\ref{s.s.time}). In particular,  
    \begin{equation}
        T\in[T_{min},T_{max}]~, 
    \end{equation}
    as apparatuses allowing only windows shorter than $T_{\min}$ fail to provide sufficient time for any observer to perform measurements, and those offering windows longer than $T_{\max}$ are redundant, as no observer requires durations beyond some defined threshold;
    \item the available energy scale, parametrised by the number $N$ of the apparatus' components (see Sec.~\ref{s.s.energy}). In particular,
    \begin{equation}
        N\in[N_{min},N_{max}]~,
    \end{equation}
    as apparatuses smaller than $N_{\min}$ are rendered nonfunctional due to susceptibility to external noise, while those exceeding $N_{\max}$ cannot be activated with some reasonable maximal available energy;
\end{itemize}

A notion is quality is then obtained by observing that a large value of $T$ means the apparatus allows a wider selection of measurement intervals with longer durations, and a small value of $N$ means the apparatus is more cost-effective in terms of energy usage. As it is clear, apparatuses providing longer windows at a smaller energy cost are generally better than those providing smaller windows at a higher energy cost. This classification is visually depicted in Fig.~\ref{f.plot}, where apparatuses in the top-left corner of the region 
\begin{equation}
    \mathcal{R}=[N_{min},N_{max}]\times[T_{min},T_{max}]
\end{equation}
are better than those in the bottom-right corner. 

Given this qualitative definition of better and worse measurement apparatuses, we can determine the best choice of couplings by studying how $\mathcal{A}(t)$ varies depending on $g_k$ and $N$, as the overlap of the pointer states is directly related to the probability of getting a good outcome via the size of $\mathbb{T}^\epsilon_{WPRC}$ (see Sec.~\ref{s.WPRC_&_Energy}). Therefore, we study $\mathcal{A}(t)$ to determine which $g$ makes better $g$-ordered apparatuses, which $I$ makes better $G_I$-disordered apparatuses, and compare these to establish which choice is the most suitable for a given observer. 

\subsubsection{$g$-ordered apparatuses}

As anticipated, we here study $\mathcal{A}(t)$ to determine which $g$ makes the best $g$-ordered apparatus within the region $\mathcal{R}$. In this case, $\mathcal{A}(t)$ is periodic with period $\pi/2g$ and has one minimum for each periodicity interval, the first being in $\pi/4g$.
If the initial state of at least one of the qubits of the apparatus is s.t. $|\alpha_k|=1/\sqrt{2}$, then 
\begin{equation}
\mathbb{T}_{\rm{PRC}}=\left\lbrace t\in\mathbb{R}^+~|~t=\frac{\pi}{4g}+\frac{n\pi}{2g}~,~ n\in\mathbb{N}\right\rbrace ~;
\end{equation}
otherwise
\begin{equation}
\mathbb{T}_{\rm{PRC}}=\emptyset~.
\end{equation}
In both cases, there is no time interval an observer can use to do measurements, meaning we must resort to the WPRC.

As illustrated in Fig.~\ref{f.ordered}, $\mathcal{A}(t)$ approaches values closer to zero within the periodicity interval for increasing $N$; on the contrary, it is equal to one at $t_n=\pi n/g,~n\in\mathbb{N}$, regardless of $N$. Therefore, the time ranges available for measurements are
\begin{equation}
\mathbb{T}^{\epsilon}_{\rm{WPRC}}=\bigcup_{k\in\mathbb{N}} I^{(o)}_k~,
\end{equation}
with
\begin{equation}
    I^{(o)}_k=\left[\frac{k\pi}{2g}-\tau_\epsilon,\frac{(k+1)\pi}{2g}+\tau_\epsilon\right]~,
\end{equation}
where $\tau_\epsilon$ is the smallest positive time satisfying $\tau_\epsilon=\mathcal{A}^{-1}(\epsilon)$; Notice that $\tau_\epsilon$ is a decreasing function of with $N$ and $\epsilon$, and an increasing function of $g$. Therefore, $g$ decreases\footnote{Notice that this also means that they require an observer to wait longer times to realise the optimal measurement window required by Eq.~\eqref{e.good_or_perfect}.} when approaching the top-left corner of the classification diagram in Fig.~\ref{f.plot} (green reagion).

\subsubsection{$G_I$-disordered apparatus}
Following the logic above, we here study $\mathcal{A}(t)$ to determine which $I$ makes the best $G_I$-disordered apparatus within $\mathcal{R}$. In this case, $\mathcal{A}(t)$ is either periodic or quasi-periodic and generally has many minima within each (quasi-)period. For any finite $N$, $\mathbb{T}_{\rm{PRC}}$ is a discrete collection of points, and $\mathbb{T}^{\epsilon}_{\rm{WPRC}}$ is a collection of intervals of various lengths. At sufficiently large values of $N$, any interval $\mathbb{T}$ contains more points of $\mathbb{T}_{\rm{PRC}}$ than the corresponding construction in the ordered case. However, since $\mathbb{T}_{\rm{PRC}}$ has measure zero, this fact is not sufficient to classify the disordered case as better than the ordered one: again, no interval can be used to do measurements, and one should resort to the WPRC.

\begin{figure}
    \centering
    \includegraphics[width=0.45\textwidth]{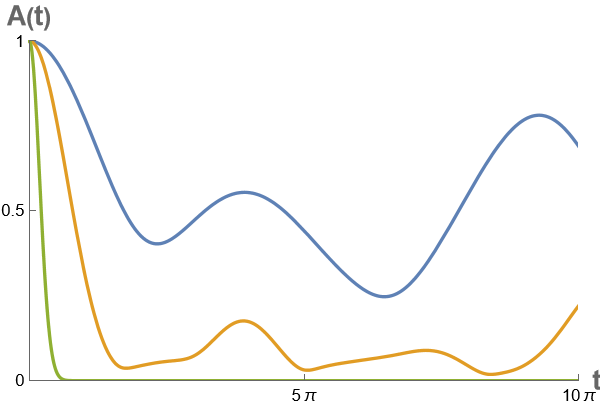}
    \caption{Availability $\mathcal{A}(t)$ for a $G_I$-disordered apparatus with $I=[0,0.2]$, for $N=5$ (blue), $N=10$ (orange), and $N=100$ (green).\vspace{2.4em}}
    \label{f.disordered}
\end{figure}

As illustrated in Fig~\ref{f.disordered}, $\mathcal{A}(t)$ is squished towards zero for increasing $N$, as in the ordered case. However, since the couplings are chosen randomly, the (quasi-)periodicity of $\mathcal{A}(t)$ is longer (and the points for which it evaluates one are rarer) than in the ordered case. Therefore, for $N$ sufficiently large, the intervals composing $\mathbb{T}^{\epsilon}_{\rm{WPRC}}$ become very long and with few scattered interruptions. Specifically,
\begin{equation}
\mathbb{T}^{\epsilon}_{\rm{WPRC}}=\bigcup_{k\in\mathbb{N}}I^{(d)}_k
\end{equation}
with 
\begin{equation}
     I_k^{(d)}=\left[ t_{k}+\tau_k(N,g_k,\epsilon),t_{k+1}-\tau'_k(N,g_k,\epsilon) \right]~,
     \label{e.I^d}
\end{equation}
where $t_k$ are the sparse ones of $\mathcal{A}(t)$. Notice that all quantities appearing in this case, namely $t_k$, $\tau_k(N,g_k,\epsilon)$ and $\tau'_k(N,g_k,\epsilon)$, are extremely hard to identify analytically even when the set of random couplings is given.

To compare different $G_I$-disordered apparatuses, it is important to characterise the role played by $I$ in determining the shape of $\mathcal{A}(t)$. First, $I$ must be large enough to give a non-negligible contribution: if the region is too small, at short times, $\mathcal{A}(t)$ will look the same as in the ordered case. To see this, suppose that $I=[g-\delta g, g+\delta g]$, for a small enough $\delta g$. Then, for example, Eq.~\eqref{e.selected delta} becomes
\begin{equation}
    \Delta(t)\simeq\cos(2g t)^{N}\left[1-\frac{2t}{\cos{2gt}}\sum_k\delta g_k\sin(2g_k t)\right]~,
\end{equation}
where $g_k=g+\delta g_k$ with $|\delta g_k|<\delta g$. As it is clear, the function cannot be distinguished from its $g$-ordered version at short times (i.e. $t\ll1/2g_k$, $\forall k$). Of course, the contribution of the randomness will appear later, yet this might be too late for an observer unaware of the sampling region size to take action. Second, while the above argument says that big sampling regions provide better results, smaller $g_k$ generally give better results, as in the $g$-ordered case. Therefore, a good sampling region is wide, yet only contains small values of $g$. This means that apparatuses in the top-left corner of the diagram in Fig.~\ref{f.plot} (green region) are characterised by wide intervals centred around some small $g$.

\subsubsection{Interlude: large-$N$ apparatuses}
By looking at Eq.~\eqref{cprod}, we get that
\begin{equation}
    \lim_{N\to\infty}\tau_\epsilon=0~,
\end{equation}
and 
\begin{equation}
    \lim_{N\to\infty}\tau_k(N,g_k,\epsilon)=0=\lim_{N\to\infty}\tau'_k(N,g_k,\epsilon)~;
\end{equation}
these equations mean that, in the limit, the WPRC intervals cover the real line except for some discrete collection of points ($t_n=\pi n/g,~n\in\mathbb{N}$ in the $g$-ordered case, sparse $t_k$ in the $G_I$-disordered one) and the apparatus becomes perfectly reliable. This result is robust under variations of the $g_k$'s (both in the ordered and disordered case) and consistent with previous literature on the quantum-to-classical crossover. As shown in Ref.~\cite{FotiEtAl19}, any environment behaves as a (perfectly reliable) measurement apparatus in the large-$N$ limit. However, these apparatuses always fall outside $\mathcal{R}$, meaning that no observer has sufficient energy to switch on the interaction: large-$N$ apparatuses are inaccessible in the sense of Sec.~\ref{s.s.energy}.

\subsubsection{Order makes a predictable apparatus, disorder makes an accessible one}
In this section, we use the results of above to compare $g$-ordered and $G_I$-disordered apparatuses and discuss what choice of coupling coefficients is better for an observer characterised by the point $(N_0,T_0)$ in the $N$-$T$ diagram (see Fig.~\ref{f.plot}).

Before starting, let us notice that there are (at least) a $g$-ordered and a $G_I$-disordered apparatus for each point of the $N$-$T$ diagram, meaning the observer can choose either and be able to perform measurements. Therefore, our discussion must be based on grounds other than the possibility of using one or the other. Indeed, we will again use the shape of $\mathcal{A}(t)$ to discuss the practical advantages of the various options and provide a rule for choosing based on the observer's specific needs.

On the one hand, comparing a $g$-ordered and a $G_I$-disordered apparatus such that $g\in I$ shows that, most often,
\begin{equation}
    t_{k+1}-t_k\gg \frac{\pi}{2g}
    \label{e.compare}
\end{equation}
where the $t_k$ are those defining $I_k^{(d)}$ in Eq.~\eqref{e.I^d}, meaning that the disordered apparatus generally provides longer measurement windows than the ordered one. Moreover, since, at finite $N$,
\begin{equation}
    I_k^{(o)}\subset \left[\frac{k\pi}{2g},\frac{(k+1)\pi}{2g}\right]~,
\end{equation}
a $g$-ordered apparatus cannot generate measurement windows longer than $\pi/2g$, meaning that if we need longer windows, we must change the coupling or utilize a disordered apparatus. Hence, disordered apparatuses give longer measurement windows at lower energy costs.


On the other hand, for the disordered case to be reliable, one must choose $N$ large enough to dampen the random spikes appearing in $\mathcal{A}(t)$ or deal with the possibility of spurious non-accurate measurement (meaning that they don't satisfy the WPRC). For this reason, a disordered measurement apparatus is less predictable than an ordered one at any finite value of $N$.

Considering all these factors together, we propose that:
\begin{itemize}
    \item disordered apparatuses are better if the observer values energy saving over reliability. This is because they provide long measurement windows at low energy cost, yet they may give wrong results at unpredictable times due to the random spikes of $\mathcal{A}(t)$. Disordered apparatuses are more \textit{accessible} than ordered ones.
    \item ordered apparatuses provide a more reliable behaviour if one knows the interaction coupling and can spend a lot of energy. This is because $\mathcal{A}(t)$ can be computed analytically but, to get long measurement windows, one needs large apparatuses. Ordered apparatuses are more \textit{reliable} than disordered ones.
\end{itemize} 

Therefore, the observer willing to save energy should prefer disordered apparatuses, while careful experimenters should favour ordered ones.

\section{Conclusions}
\label{s.conclusions}
Assuming von Neumann's premeasurements and the Ozawa scheme, we introduced the notion of Weak Probability Reproducibility Condition (WPRC), hence solving the problem of having no time intervals to perform measurements on an apparatus unitarily coupled to a measured system. Then, we proposed physically meaningful time and energy constraints an observer should require over a measurement apparatus's dynamics to be utilisable: as we focus on observer-dependent constraints, our approach is fundamentally operational. In this way, we found a classification of measuring apparatuses based on their reliability (i.e. their ability to faithfully represent a system's populations on their pointer basis in a given time window) and accessibility (i.e. their ability to extract information from the target system using a reasonably small amount of energy). These qualities are directly related to the time behaviour of the pointer basis overlap, here called availability. Then, by analysing the paradigmatic example of $N$ qubits measuring a target qubit, we found that measurement apparatuses with randomly selected \textit{disordered} couplings are versatile and energy-efficient, hence having high availability. Conversely, apparatuses characterised by uniformly distributed \textit{ordered} couplings exhibit higher levels of reliability.

Thanks to its generality and the fact that computing the reliability and accessibility of an apparatus only requires knowing its pointer states overlap, our construction provides a simple formal tool with broad applicability that can be used to classify a vast number of measurement apparatus designs. Therefore, there are several avenues for future developments stemming from our work. First, one could investigate random coupling distributions other than the one analysed herein (namely, the uniform one); examining different coupling scenarios and their impact on the reliability and accessibility of the qubit measurement apparatus might give insights into how to construct better quantum measurements in real-world scenarios and pave the way for using our classification beyond the example we explored. Second, investigating other models might illuminate the suitability of specific quantum systems as measurement apparatuses, providing valuable insights for various experimental contexts. 
Finally, when delving into practical applications of our classification, one should also consider the additional cost associated with the computational effort required to identify the pointer basis ${\ket{\Xi^{\pm}(t)}}$. Roughly speaking, the computational complexity of this task scales linearly with the size of the apparatus in both the disordered and ordered scenarios, as one needs to store $O(3N)$ variables for the former and $O(2N)$ variables for the latter (see the model setup in Sec.~\ref{s.example} for further details). When translating our theoretical framework into practical use, one should further investigate this additional cost via the tools of classical complexity theory.


\section*{Acknowledgments}
We thank Marco Cattaneo for helpful comments. N.P. acknowledges financial support from the Magnus Ehrnrooth Foundation and the Academy of Finland via the Centre of Excellence program (Project No. 336810 and Project No. 336814). P.V. acknowledges financial support from PNRR MUR project PE0000023-NQSTI and declares to have worked in the framework of the Convenzione Operativa between the Institute for Complex Systems of CNR and the Department of Physics and Astronomy of the University of Florence.

\bibliography{bib}

\newpage
\onecolumngrid

\appendix
\section{PRC times provide more information than WPRC intervals}
\label{s.mutual information}
In this appendix, we show how using pointers satisfying the WPRC in place of the PRC affects the amount of information about the system an observer can extract from the apparatus. To this end, we first study how Ozawa's dynamic increases the information the environment has about the system. The key quantity to understanding how much information can be extracted from one system regarding another is called mutual information, defined by
\begin{equation}
\mathcal{I}(\Gamma : \Xi)=S(\Gamma)+S(\Xi)-S(\psi)~,
\end{equation}
where $S(A)=-\text{Tr}(\rho_A \log\rho_A)$ is the von Neumann entropy of the system $A$. At $t=0$, the system is in a separable state: the total density operator is a tensor product of local projectors onto the state vectors of the individual subsystem, meaning that
\begin{equation}
S(\Gamma)=S(\Xi)=S(\psi)=0~,
\end{equation}
and hence
\begin{equation}
    \mathcal{I}(\Gamma : \Xi)_{t=0}=0~.
\end{equation}
The interpretation of the above expression is clear: in the initial state, the two systems share no information. Indeed, this was already clear from the possibility of factoring the total density operators into a tensor product of pure states related to the subsystems. On the contrary, once a value of $t$ in the set $\mathbb{T}_{\rm{PRC}}$ is chosen, the reduced density operators $\rho_\Gamma$ and $\rho_\Xi$ have entropy
\begin{equation}
S(\Gamma)=S(\Xi)=-\sum_{\gamma}|c_{\gamma}|^2\log |c_{\gamma}|^2~,
\end{equation} 
from which we get
\begin{equation}
\mathcal{I}(\Gamma : \Xi)_{t\in \mathbb{T}_{\rm{PRC}}}=-2\sum_{\gamma}|c_{\gamma}|^2\log |c_{\gamma}|^2~.
\label{e.entropy}
\end{equation}
Therefore, at the times in $\mathbb{T}_{\rm{PRC}}$ during the premeasurement process, the information shared between the two systems increases by
\begin{equation}
\Delta\mathcal{I}(\Gamma : \Xi)=\mathcal{I}(\Gamma : \Xi)_{t\in \mathbb{T}_{\rm{PRC}}}-\mathcal{I}(\Gamma : \Xi)_{t=0}=-2\sum_{\gamma}|c_{\gamma}|^2\log |c_{\gamma}|^2\geq 0~.
\end{equation}
In particular, in this case, the extracted information is maximal. However, replacing the PRC with the WPRC reduces the information one can extract about the system from the apparatus. For simplicity, we slightly relax the PRC by requiring that
\begin{equation}
\langle\Xi^{\gamma}\ket{\Xi^{\gamma'}}=\begin{dcases}
    \delta_{\gamma,\gamma'}~&\forall \gamma,\gamma'\neq 1,2\\
    1~&\rm{if}~\gamma=\gamma'=1,2\\
    \epsilon~&\rm{otherwise}
\end{dcases}~.
\end{equation}
In this case, the density operator is expressed as
\begin{equation}
\rho_{\gamma}=\begin{pmatrix} \Delta_{+} & 0 & 0 & \\ 0 & \Delta_{-} & 0 & \cdots \\ 0 & 0 & |c_3|^2 &  \\ &\vdots& &\ddots
\end{pmatrix}
\end{equation}
in the basis over which it is diagonal,
where
\begin{equation}
\Delta_{\pm}=\frac{|c_1|^2+|c_2|^2}{2}\pm\frac{\sqrt{(|c_1|^2-|c_2|^2)^2+4\epsilon^2}}{2}
\end{equation}
Therefore, for $\epsilon\ll 1$, we get
\begin{equation}
\begin{dcases}
    \Delta_+\simeq |c_1|^2+\frac{\epsilon^2}{|c_1|^2-|c_2|^2}\\
    \Delta_-\simeq |c_2|^2-\frac{\epsilon^2}{|c_1|^2-|c_2|^2}
\end{dcases}~~~.
\end{equation}
In  this case, the entanglement entropy
\begin{equation}
S^{\epsilon\neq 0}(\Gamma)=-\sum_{\gamma\geq 3}|c_{\gamma}|^2\ln|c_{\gamma}|^2-\Delta_+\ln\Delta_+-\Delta_-\ln\Delta_-~,
\label{e.entropy_WPRC}
\end{equation}
can be compared with Eq.~\eqref{e.entropy} to obtain
\begin{equation}
S^{\epsilon\neq 0}(\Gamma)\simeq S^{\epsilon=0}(\Gamma)-\epsilon^2\frac{\ln{|c_1|^2}-\ln{|c_2|^2}}{|c_1|^2-|c_2|^2}=S^{\epsilon=0}(\Gamma)-A
\end{equation}
where $A>0$. Hence, using Eq.~\eqref{e.entropy_WPRC} to compute the mutual information gives
\begin{equation}
\Delta\mathcal{I}^{\epsilon\neq 0}=\Delta\mathcal{I}^{\epsilon= 0}-2A~~\Rightarrow ~~\Delta\mathcal{I}^{\epsilon= 0}>\Delta\mathcal{I}^{\epsilon\neq 0}
\end{equation}
meaning that the information extracted from the premeaurement process at the times for which the WPRC is valid is smaller than that obtained under PRC.

\end{document}